\documentclass[10pt,final]{IEEEtran}
\usepackage{graphicx}
\usepackage{cite}
\usepackage{color}
\usepackage[cmex10]{amsmath}
\begin{document}
%
\title{Queueing Analysis of Unicast IPTV With User Mobility and Adaptive Modulation and Coding in Wireless Cellular Networks}

\author{\IEEEauthorblockN{Mingfu Li} \\
\IEEEauthorblockA{\normalsize Department of Electrical Engineering, School of Electrical and Computer Engineering, \\ College of Engineering, Chang Gung University, Tao-Yuan, 33302 Taiwan\\ 
Email: mfli@mail.cgu.edu.tw, Phone: +886-3-2118800\#5676}}


\maketitle

\begin{abstract}
Unicast IPTV services that can support live TV, video-on-demand (VoD), video conferencing, and online gaming applications over broadband wireless cellular networks have been becoming popular in recent years. However, video streaming services significantly impact the performance of wireless cellular networks because they are bandwidth hogs. To maintain the system performance, effective admission  control and resource allocation mechanisms based on an accurate mathematical analysis are required. On the other hand, the quality of a wireless link usually changes with time due to the user mobility or time-varying channel characteristics. To counteract such time-varying channels and improve the spectral efficiency, adaptive modulation and coding (AMC) scheme can be adopted in offering unicast IPTV services for mobile users. In this paper, closed-form solutions for the bandwidth usage, blocking rate, and dropping rate of unicast IPTV services over wireless cellular networks were derived based on the novel queueing model that considers both user mobility and AMC.  Simulations were also conducted to validate the accuracy of analytical results. Numerical results demonstrate that the presented analytical results are accurate. Based on the accurate closed-form solutions, network providers can implement precise admission control and resource allocation for their networks to enhance the system performance.
\end{abstract}

\begin{IEEEkeywords}
Adaptive modulation and coding (AMC), bandwidth usage, blocking rate, dropping rate, mobile IPTV, unicast, user mobility, wireless cellular networks.
\end{IEEEkeywords}

\IEEEpeerreviewmaketitle

\section{Introduction}
\IEEEPARstart{I}{nternet} Protocol Television (IPTV) services have been becoming increasingly popular among telecommunication companies \cite{0,0_1}. Recently, due to the significant growth in the amount of mobile devices, access bandwidth over wireless networks, and advanced video coding  technologies \cite{0_2,d,d1,7}, mobile IPTV \cite{a,b,c,c_1} has
become viable over wireless cellular networks. Several
works \cite{d,d1,7,a,b,c,c_1,c_2} have focused on the architecture and standardization for providing mobile IPTV services. According to a recent study by Cisco \cite{0_3}, mobile video traffic exceeded 50\% of mobile data traffic for the first time in 2012 and was 53\% of traffic by the end of 2013.  Mobile video will grow at a compound annual growth rate (CAGR) of 69\% between 2013 and 2018, indicating that it is having an immediate impact on traffic today and in the future. Most of the video traffic is from live TV and video-on-demand (VoD) applications, and VoD traffic will nearly triple by 2017. Therefore, the impact of video services on the performance of mobile networks is significant and must be investigated.


In wireless cellular networks, the channel conditions are time-varying because of multipath fading, radio interference, and user mobility. To provide robust transmissions and improve the spectral efficiency, the adaptive modulation and coding (AMC) scheme for wireless link adaption was proposed  and analyzed \cite{AMC1, AMC2, AMC3,8,9,Mingfu1}. The AMC scheme can dynamically adjust the modulation and coding scheme (MCS) of a wireless link according to its signal-to-noise ratio (SNR), signal-to-interference-plus-noise ratio (SINR),  or channel quality indicator (CQI). When the wireless link has a higher SNR, SINR, or CQI, the MCS with a higher data rate can be used to improve the spectral efficiency. AMC has been widely used in wireless communication networks, such as WiMAX \cite{AMC1}, Enhanced Data rates for GSM Evolution (EDGE) \cite{AMC2} and High-Speed Downlink Packet Access (HSDPA) \cite{AMC3}, for link adaptation. A closed-form expression for the spectral efficiency in single-carrier frequency-division multiple access (SC-FDMA) when AMC is applied was derived in \cite{8}. In \cite{9}, the packet drop rate, packet average delay and throughput were derived based on the Markovian queueing model when AMC is used.  The blocking rate and spectrum consumption of multicast IPTV services were studied by simulations in \cite{Mingfu1}. All results in \cite{8,Mingfu1,9} show that AMC can significantly improve the spectral efficiency and network performance.

Three basic transport technologies for video streaming applications include unicast, broadcast, and multicast. Broadcast and multicast schemes can provide a good bandwidth gain at the heavy load condition \cite{Mingfu2,5}, but the implementation and management of AMC in them for improving the spectral efficiency are complicated and difficult. For example, the issue of CQI decision and maintenance for each multicast group and the CQI feedback implosion problem \cite{10,CQI} still remain to be solved.  Unicast technology provides one-to-one connection between the base station and the mobile user. Although the bandwidth gain of unicast scheme is low, it can support both interactive VoD and live TV services with AMC link adaptation for increasing the spectral efficiency. Additionally, users prefer to watch video on-demand, rather than following a fixed schedule, to meet the anytime and anywhere requirement. Therefore, unicast is still a popular and important transport scheme for mobile video streaming applications today. 

Unicast IPTV services are bandwidth hogs network applications and have a significant impact on the performance of wireless cellular networks. Thus, to guarantee the network QoS (Quality of Services) for subscribers, effective admission control, frequency planning, and resource allocation for unicast IPTV services are extremely required. To achieve these goals, accurate performance analysis for unicast IPTV services must be provided first.  However, most of the performance analyses for IPTV in wireless networks focus on the packet-level performance, such as the packet delay/jitter, packet loss rate, and video quality/distortion. Such conventional packet-level rather than system-level performance analyses cannot resolve the resource allocation issue. 

Only few papers  \cite{AMC1,Mingfu2,3,NVoD,4} focused on the system-level performance analysis for unicast IPTV services. The work \cite{AMC1} developed a model to estimate the amount of WiMAX cell resources required for a video channel delivery, as a function of the number of video channels offered and the number of users on the cell. The paper \cite{Mingfu2} analyzed and compared the performance of unicast, multicast and broadcast IPTV services. Another paper \cite{3} analyzed the blocking probability of unicast  IPTV services as a function of the average number of users. The authors in \cite{NVoD} calculated the blocking probability of unicast TV services by Erlang's loss formula with the M/M/C/C queueing model. In \cite{4}, the performance of unicast TV services was also assessed based on the Markovian queueing model. Among these works, only the analysis models in \cite{AMC1} and \cite{4} consider the effect of MCSs on the system performance. All these presented analysis models for unicast IPTV services did not take user mobility into consideration. However, the user behavior such as the user mobility and the video watching time of a user can significantly affect the performance of networks supporting IPTV services. Hence, in this work the blocking rate, dropping rate, and bandwidth usage of unicast IPTV services are analyzed based on the queueing model that considers both AMC and user mobility. The differences of analysis methods between this work and \cite{AMC1,4} are addressed as follows. In \cite{AMC1}, the analytical model is not based on the queueing model and assumes that the number of users requesting TV channels is known. In the work \cite{4}, the authors mentioned that their proposed queueing model can be solved by constructing the generating matrix and solving the corresponding balance equations. However, it is difficult to obtain the solutions using the generating matrix since the size of the generating matrix is large. Therefore, they resorted to the approximate insensitive solutions to obtain the network performance. 

In this work, multi-dimensional system states are defined for the queueing model that considers both AMC and user mobility. The difficulties of analyzing the proposed problem include the modeling of states, formulation of global balance equations, and verification of the availability of local balance equations. First, the queueing system with multi-dimensional states is rarely studied and found in the literature. Second, the global balance equations with multi-dimensional states are not easy to be formulated. If the global balance equations cannot be formulated, then it is impossible to show whether the local balance equations exist or not. When local balance equations do not exist, conventional methods to solve the queueing system can only appeal to the transition rate (intensity) matrix $Q$ and obtain the stationary state probability vector $\vec{\pi}$ by solving $\vec{\pi}Q=\vec{0}$. However, it is very difficult to construct the matrix $Q$ and solve  $\vec{\pi}Q=\vec{0}$ because the size of the transition rate matrix $Q$ for the queueing system with multi-dimensional states is usually very large \cite{4}. It is also impossible to derive the closed-form solution by solving $\vec{\pi}Q=\vec{0}$. To overcome  this problem, the unit vector  $\vec{E_i }$ is first introduced and then the precise state transition diagram is constructed in this paper. Next, the global balance equations are formulated based on the state transition diagram and the availability of local balance equations is verified. For the queueing model with user mobility, exact local balance equations do not exist. Thus, approximate local balance equations are derived in this work. Finally, the closed-form solution of the stationary state probability is derived using the local balance equations. 
 
The novelty and contributions of this work are summarized as follows. Firstly, the user request arrival process, video watching time, user mobility, AMC, and limited time-frequency resources are simultaneously considered in the queueing analysis. The proposed analysis approach for unicast IPTV services is completely new. Secondly, the closed-form solutions, which are never found in literature, for the blocking rate, dropping rate, and bandwidth usage of unicast IPTV services are derived. Thirdly, different mobility models are simulated to validate the accuracy of our analysis. Finally, using the closed-form solutions, precise admission control and dynamic resource allocation can be implemented very efficiently.

The rest of this paper is organized as follows. Section II describes the system model. Section III analyzes the performance of unicast IPTV services over wireless cellular networks. Numerical results and performance comparisons are presented in Section IV. 
Finally, the concluding remarks are made in Section V.



\section{System Model}
Since mobile IPTV services require extra large bandwidth compared with voice and data services \cite{AMC1,4}, the performance of mobile networks is significantly dominated by mobile IPTV traffic. To guarantee the QoS of traditional voice/data services, the dedicated time-frequency resources for voice/data traffic and mobile IPTV traffic can be completely isolated \cite{3,NVoD}. Thus, in this paper only the performance of unicast IPTV services is analyzed.

\begin{figure}
\centering
\includegraphics[width=1.8in]{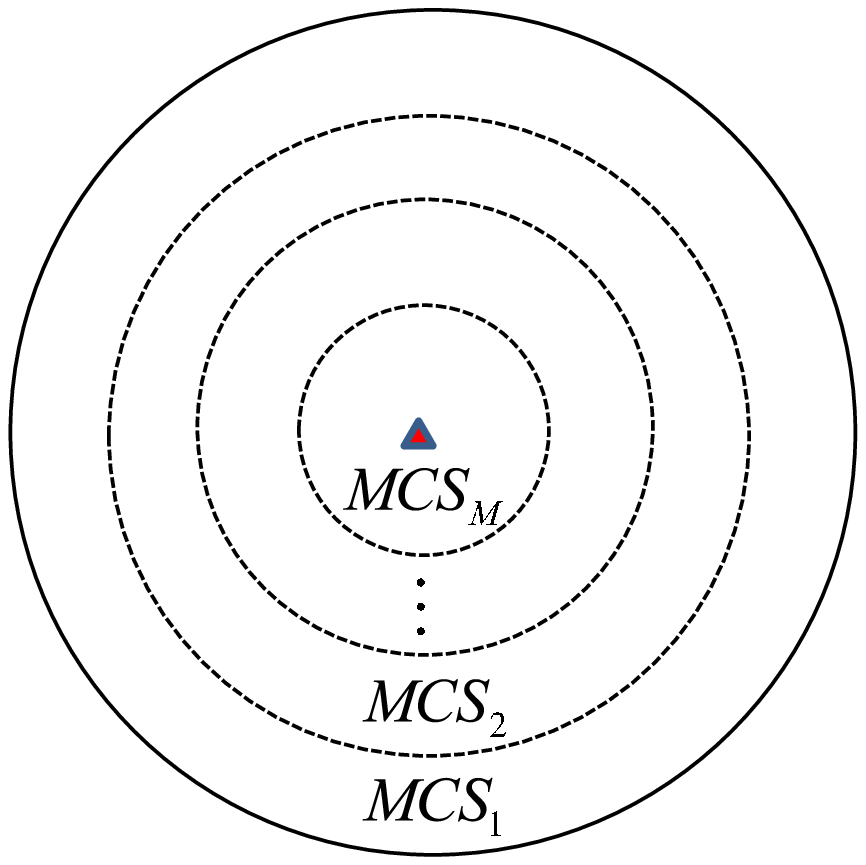}
\caption{Partitions of a cell based on the used MCSs.}\label{fig:region}
\end{figure}

Assume $M$ MCSs with respective data rates $r_1$, $r_2$, $\cdots$, $r_M$ are used in the cellular mobile network. If the signal strength attenuation is mainly due to the free space path loss, then the adaptive MCS of each unicast connection is determined according to the user's distance to the base station. Consequently, in this paper a cell is divided into $M$ concentric zones: Zones 1, 2, $\cdots$, $M$, as shown in Fig. \ref{fig:region}.  The area fraction, relative to the whole cell, of Zone $m$ is denoted by $\sigma_m$. The unicast connection for the user in Zone $m$ uses the MCS at the data rate $r_m$. The MCS rates satisfy $r_1<r_2< \cdots <r_M$.

To understand the effects of unicast IPTV services on the performance of mobile networks, the following queueing model is presented and analyzed. The considered queueing system in this paper is limited to within a cell of the mobile network since the network operation of each cell is similar. The terminology {\it bandwidth resources} is used to represent the time-frequency resources in the wireless cellular network hereafter. The amount of bandwidth resources consumed by each unicast IPTV connection, namely the bandwidth usage per unicast IPTV connection, is assumed to be $c_m$ if the MCS rate $r_m$ is used. Since the MCS rates satisfy $r_1<r_2< \cdots <r_M$, one can conclude that $c_1>c_2>\cdots>c_M$. Notably, in the proposed model the unicast IPTV streams can be constant bit rate (CBR) or variable bit rate (VBR) traffic. However, the bandwidth allocated for each unicast IPTV connection must not be less than its average bit rate in a stable system. In addition, the average data bit rate of each unicast IPTV stream, such as 128 kbps, is assumed to be the same. The total amount of bandwidth resources reserved for unicast IPTV services in each cell is assumed to be $K$. The arrival process of unicast IPTV requests is presumed to follow the Poisson process \cite{poisson} with total arrival rate $\lambda$.  The users are uniformly located in the cell. Thus, the average arrival rate of unicast IPTV requests in Zone $m$ can be denoted by $\lambda_m=\lambda \sigma_m$. The watching time of each IPTV request is exponentially distributed with mean $1/\mu$. 

To model the user mobility, the sojourn time of a user in each zone is also assumed to be exponentially distributed, i.e., a Markovian mobility model is assumed \cite{mobility}. However, the users in Zone $m$ can only move to the adjacent zones, namely Zone $m-1$ or $m+1$. That is, the transition rate of each user moving from Zone $i$ to Zone $j$, denoted by $v_{ij}$, equals zero if $|i-j|>1$. When a user moves from Zone $i$ to Zone $j$, the used MCS rate of the connection for the user must be changed from $r_i$ to $r_j$. Notably, $v_{ij}=0$ if $i=M$ and $j=M+1$. Additionally, a user in Zone 1 moving outward to the adjacent cells indicates a handover occurrence.  


\section{Queueing Analysis for Unicast IPTV Services}
\subsection{Exact Solutions for Scenario Without Change of MCS}
\begin{figure}
\centering
\includegraphics[width=2.2in]{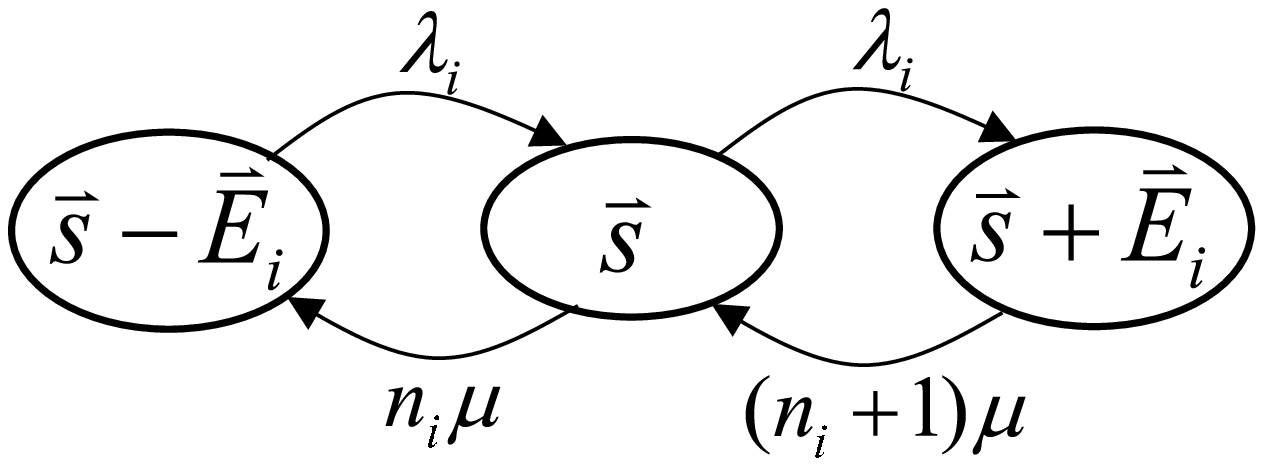}
\caption{State transition diagram without change of MCS, where $1\leq i\leq M$ and $n_i>0$, $Y\leq K-c_i$.}\label{fig1}
\end{figure}

First, the case that each user does not move to other zones during its session, i.e., $v_{ij}=0$, is considered. In such a case, all connections are not required to change their MCSs during their sessions. The system state $\vec{s}$ is defined to be the $M$-tuple vector $\vec{s}=(n_1,n_2,\cdots,n_M)$, where $n_i$ represents the number of connections using the MCS rate $r_i$. The stationary state probability of $\vec{s}$ is denoted by $p(\vec{s})$. The state transition diagram under the Markovian model is depicted in Fig. \ref{fig1}, where $\vec{E_i}$ = (0, $\cdots$, 0, $e_i=1$, 0, $\cdots$, 0) is the $M$-tuple vector that all elements are zero except the $i$-th element $e_i=1$. Since the amount of consumed bandwidth resources by  all unicast IPTV connections,  $Y=\sum_{i=1}^M n_ic_i$, must not exceed the total amount of bandwidth resources $K$, the sample space of system states is defined as $S=\{$($n_1$, $n_2$, $\cdots$, $n_M$)$|  Y=\sum_{i=1}^M n_ic_i\leq K\}$. In Fig. \ref{fig1}, if $n_i=0$, then the state $\vec{s}-\vec{E_i}$ disappears. If the bandwidth usage $Y$ in the state $\vec{s}$ is larger than $K-c_i$, then the state $\vec{s}+\vec{E_i}$ disappears. Notably, the proposed queueing model for unicast IPTV services is a connection-based model rather than a packet-based one.  The well-known classic queueing models cannot be directly applied to solve the presented queueing problem. Therefore, the closed-form solutions presented in this paper are completely novel.

When the system is in state $\vec{s}$ and the bandwidth usage $Y$ satisfies $Y\leq K-c_1$, then no user request can be rejected. Hence, according to the state transition diagram in Fig. \ref{fig1}, the global balance equation under the condition $Y\leq K-c_1$ can be expressed as follows.
\begin{eqnarray}
p(\vec{s})\sum_{i=1}^M\Big[n_i\mu I(n_i)+\lambda_i\Big] \hspace*{3.2cm}&& \nonumber \\
=\sum_{i=1}^M \Big[\lambda_i p(\vec{s}-\vec{E_i}) I(n_i)+(n_i+1)\mu p(\vec{s}+\vec{E_i})\Big], &&\label{eqn:g1}
\end{eqnarray}
where $I(n_i)$ is the indicator function defined by
\begin{equation}
I(n_i)=\left\{
\begin{array}{ll}
1, & {\rm if}\ n_i>0, \\
0, & {\rm otherwise.}
\end{array}\right.
\label{eqn:I}
\end{equation}
In (\ref{eqn:g1}), the term $n_i\mu$  must be multiplied by $I(n_i)$ because when $n_i=0$, no connection uses the MCS rate $r_i$ and the term $n_i\mu$ must disappear. The term $\lambda_i p(\vec{s}-\vec{E_i})$ must be multiplied by $I(n_i)$ because the state $\vec{s}-\vec{E_i}$ does not exist if $n_i=0$. 

When the system is in state $\vec{s}$ and the bandwidth usage $Y$ satisfies $K-c_m< Y\leq K-c_{m+1}$, where $1\leq m\leq M-1$, then the requests from the users located in Zone $i$, where $i\leq m$, must be rejected. Hence, the global balance equation under the condition $K-c_m< Y\leq K-c_{m+1}$ becomes
\begin{eqnarray}
p(\vec{s})\left\{\sum_{i=1}^M n_i\mu I(n_i)+\sum_{i=m+1}^M\lambda_i \right\} \hspace*{3.2cm}&& \nonumber \\
=\sum_{i=1}^M \lambda_i p(\vec{s}-\vec{E_i})I(n_i) +\sum_{i=m+1}^M (n_i+1)\mu p(\vec{s}+\vec{E_i}). &&\label{eqn:g2}
\end{eqnarray}

When the system is in state $\vec{s}$ and the bandwidth usage $Y$ satisfies $K-c_M< Y\leq K$, then all requests must be rejected. Hence, the global balance equation under the condition $K-c_M< Y\leq K$ becomes
\begin{eqnarray}
p(\vec{s})\sum_{i=1}^M n_i\mu I(n_i)&=&
\sum_{i=1}^M \lambda_i p(\vec{s}-\vec{E_i})I(n_i).  \label{eqn:g3}
\end{eqnarray}

It is difficult to solve the stationary state probability $p(\vec{s})$ using the global balance equations (\ref{eqn:g1}), (\ref{eqn:g2}), and (\ref{eqn:g3}). However, the following local balance equation
\begin{equation}
\lambda_i p(\vec{s})=(n_i+1)\mu p(\vec{s}+\vec{E_i}) \label{eqn:l1}
\end{equation}
can satisfy all global balance equations (\ref{eqn:g1}), (\ref{eqn:g2}), and (\ref{eqn:g3}). The reason is explained as follows. From (\ref{eqn:l1}), one can deduce the following equations:
\begin{equation}
\sum_{i}\lambda_i p(\vec{s})=\sum_{i} (n_i+1)\mu p(\vec{s}+\vec{E_i}), \label{eqn:l2}
\end{equation}
\begin{equation}
\sum_{i}\lambda_i p(\vec{s}-\vec{E_i}) I(n_i)=\sum_{i} n_i\mu p(\vec{s}) I(n_i). \label{eqn:l3}
\end{equation}
Obviously, both Eqs. (\ref{eqn:g1}) and (\ref{eqn:g2}) can be derived by combining (\ref{eqn:l2}) and (\ref{eqn:l3}), and Eq. (\ref{eqn:l3}) implies that (\ref{eqn:g3}) holds. 

In queueing theory, the global balance equations are a set of equations that in principle can always be solved to give the stationary state probabilities of a queueing system. In some situations, terms on either side of the global balance equations cancel. Then the global balance equations can be partitioned into a set of local balance equations. The structure of local balance equations is usually much simpler than global balance equations. 
If there exists a solution for the local balance equations, then this solution is also the 
solution of the global balance equations. Accordingly, the stationary state probability $p(\vec{s})$ can be solved using the local balance equation (\ref{eqn:l1}). For all states $\vec{s}=(n_1,\cdots,n_i,\cdots,n_M)\in S$ and $n_i>0$, the local balance equation in (\ref{eqn:l1}) can be rewritten as follows.
\begin{eqnarray}
&&\lambda_i\cdot p(n_1,\cdots,n_{i-1},n_i-1,n_{i+1},\cdots,n_M)\nonumber \\ &=&n_i\mu\cdot p(n_1,\cdots,n_{i-1},n_i,n_{i+1},\cdots,n_M).
\end{eqnarray}
Then one can iteratively derive the following result
\begin{eqnarray}
&&p(n_1,\cdots,n_{i-1},n_i,n_{i+1},\cdots,n_M) \nonumber \\ &=&\frac{(\lambda_i/\mu)^{n_i}}{n_i!}\cdot p(n_1,\cdots,n_{i-1},0,n_{i+1},\cdots,n_M).\ \ \label{eqn:gen_s1}
\end{eqnarray}
Applying the similar procedure in (\ref{eqn:gen_s1}) to other elements $n_j$'s,  the stationary state probability $p(\vec{s})$ can be derived as
\begin{equation}
p(\vec{s})=p(\vec{0}) \prod_{i=1}^M\frac{(\lambda_i/\mu)^{n_i}}{n_i!}, \label{eqn:ps}
\end{equation}
where $p(\vec{0})\equiv p(0,\cdots,0)$.

Since the sum of all stationary state probabilities must equal 1, the following equation must hold.
\begin{eqnarray}
\sum_{\vec{s}\in S}p(\vec{s}) &=& p(\vec{0})\sum_{\forall (n_1,\cdots,n_M)\in S\ } \prod_{i=1}^M\frac{(\lambda_i/\mu)^{n_i}}{n_i!}  \nonumber \\
&=&1
\end{eqnarray}
Thus,
\begin{equation}
p(\vec{0})=\frac{1}{\displaystyle\sum_{\forall (n_1,\cdots,n_M)\in S\ } \prod_{i=1}^M\frac{(\lambda_i/\mu)^{n_i}}{n_i!} }. \label{eqn:p0}
\end{equation}

Define $B_m$ to be the set of all states $\vec{s}$ satisfying $Y>K-c_m$, where $1\leq m\leq M$.  If the system is in the state $\vec{s}\in B_m$, then the requests from the users located in Zone $i$, where $i\leq m$, must be rejected. In this paper, the blocking rate is defined as the probability that a request is rejected upon its arrival owing to the system lack of bandwidth resources to serve it. Accordingly, the blocking rate $P_b$ can be computed as follows:
\begin{equation}
P_b=\sum_{m=1}^M Pr\{{\rm rejected} | {\rm Zone}\ m\}Pr\{{\rm Zone}\ m\},
\label{eqn:Pb}
\end{equation}
where $Pr\{{\rm Zone}\ m\}$ = $\sigma_m$ and $Pr\{{\rm rejected}|{\rm Zone}\ m\}$ is the conditional probability that a request is rejected upon its arrival, given that the request is from Zone $m$. Thus, $Pr\{{\rm rejected}|{\rm Zone}\ m\}$ can be expressed as follows:
\begin{eqnarray}
Pr\{{\rm rejected}|{\rm Zone}\ m\}= \frac{\displaystyle\sum_{\ \forall (n_1,\cdots,n_M)\in B_m }\prod_{i=1}^M \frac{(\lambda_i/\mu)^{n_i}}{n_i!}}{\displaystyle \sum_{\forall (n_1,\cdots,n_M)\in S\ } \prod_{i=1}^M \frac{(\lambda_i/\mu)^{n_i}}{n_i!}}.&& \label{eqn:reject}
\end{eqnarray}
Consequently, the blocking rate $P_b$ in (\ref{eqn:Pb}) can be rewritten as
\begin{equation}
P_b=\frac{\displaystyle\sum_{m=1}^M \sigma_m \sum_{\ \forall (n_1,\cdots,n_M)\in B_m } \prod_{i=1}^M \frac{(\lambda_i/\mu)^{n_i}}{n_i!}}{\displaystyle \sum_{\forall (n_1,\cdots,n_M)\in S\ } \prod_{i=1}^M \frac{(\lambda_i/\mu)^{n_i}}{n_i!}}. \label{eqn:blocking}
\end{equation}
As to the average bandwidth usage, $E(Y)$, it can be computed as follows:
\begin{equation}
E(Y)=\frac{\displaystyle\sum_{\forall (n_1,\cdots,n_M)\in S}\left( \sum_{i=1}^M n_ic_i\right) \prod_{i=1}^M\frac{(\lambda_i/\mu)^{n_i}}{n_i!}}{\displaystyle\sum_{\forall (n_1,\cdots,n_M)\in S\ } \prod_{i=1}^M\frac{(\lambda_i/\mu)^{n_i}}{n_i!}}. \label{eqn:EY}
\end{equation}

\subsection{Approximate Solutions for Scenario with Change of MCS}
\begin{figure}
\centering
\includegraphics[width=2.2in]{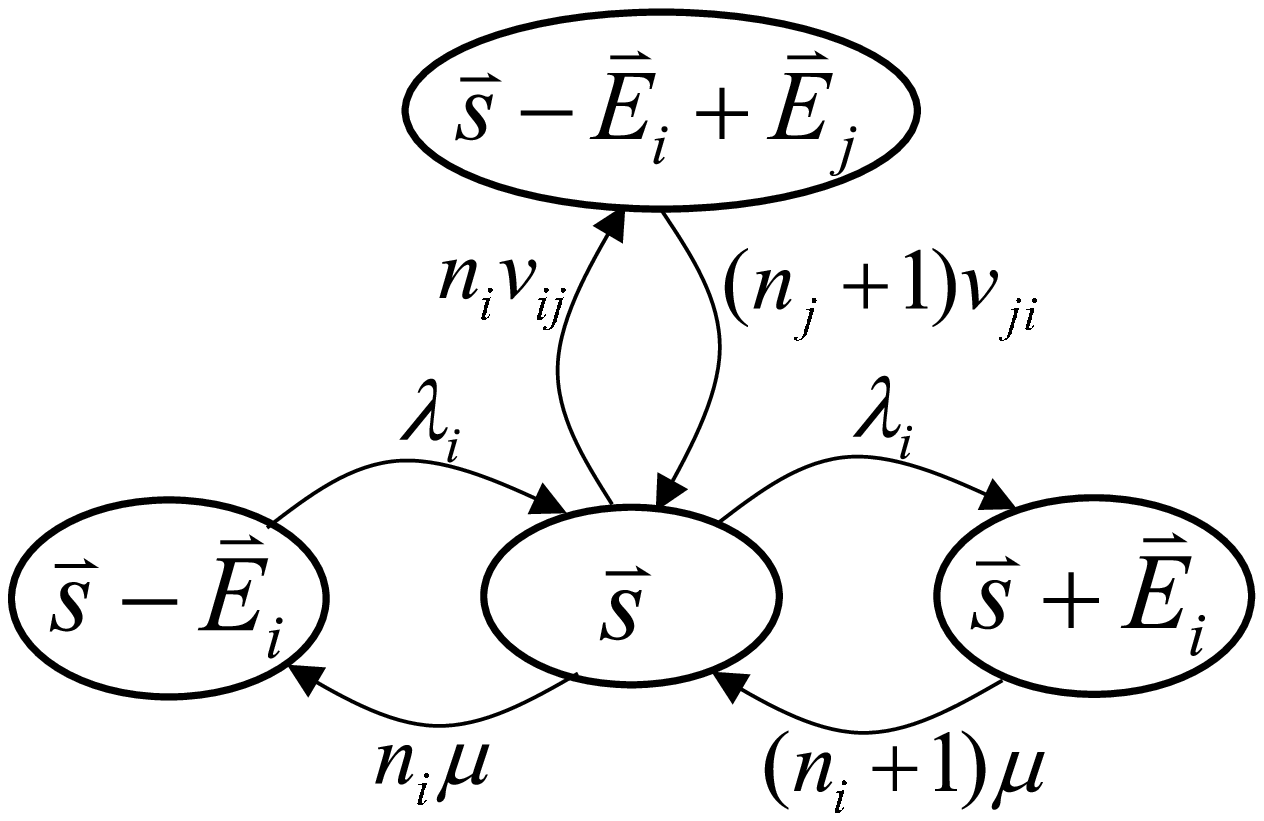}
\caption{State transition diagram with change of MCS, where $1\leq i, j\leq M$, $i\neq j$, and $n_i>0$, $Y\leq K-\max\{c_i, c_j-c_i\}$.} \label{fig:state_transition2}
\end{figure}

When the used MCS of a unicast connection may change during its session, i.e., $v_{ij}\neq 0$, the state transition diagram in Fig. \ref{fig1} must be revised to be that depicted in Fig. \ref{fig:state_transition2}, where transitions between states $\vec{s}$ and $\vec{s}-$$\vec{E_i}+$$\vec{E_j}$ become possible. When the MCS rate of a unicast connection changes from $r_i$ to  $r_j$, the state transits from $\vec{s}$ to $\vec{s}-$$\vec{E_i}+$$\vec{E_j}$. Hence, the global balance equations (\ref{eqn:g1}), (\ref{eqn:g2}), and (\ref{eqn:g3}) must be modified. For example, Eq. (\ref{eqn:g1}) must be revised into
\begin{eqnarray}
&&p(\vec{s})\sum_{i=1}^M\Bigg\{n_i\Big(\mu +\sum_{j\neq i} v_{ij}\Big) I(n_i)+\lambda_i \Bigg\}  \nonumber \\
&=&\sum_{i=1}^M \Big [\lambda_i p(\vec{s}-\vec{E_i}) I(n_i)+(n_i+1)\mu p(\vec{s}+\vec{E_i})\Big ]  \hspace*{0.5cm} \nonumber \\ &&+ \sum_{i=1}^M\sum_{j\neq i} (n_j+1) v_{ji} p(\vec{s}-\vec{E_i}+\vec{E_j}) I(n_i).  \label{eqn:gg1}
\end{eqnarray}

According to (\ref{eqn:gg1}), the stationary state probability $p(\vec{s})$ can no longer be solved using the same local balance equation (\ref{eqn:l1}) and the problem becomes complicated. To the best of our knowledge, the closed-form solution for such a queueing system is still not found in the literature. However, an approximate  closed-form solution can be obtained using the approach proposed in the following. 
According to the law of total probability \cite{prob},  $p(\vec{s}-\vec{E_i}+\vec{E_j})$ can be expressed as follows:
\begin{eqnarray}
&& p(\vec{s}-\vec{E_i}+\vec{E_j}) \nonumber \\ & = & Pr\big\{\vec{s}-\vec{E_i} +\vec{E_j}\big |\vec{s}-\vec{E_i}\big\}\cdot p(\vec{s}-\vec{E_i}) \nonumber \\ && +Pr\big\{\vec{s}- \vec{E_i}+\vec{E_j} \big | \vec{s}\big\}\cdot p(\vec{s}) \nonumber \\ && + Pr\big\{\vec{s}- \vec{E_i}+\vec{E_j}\big | \vec{s}+\vec{E_j}\big\}\cdot p(\vec{s}+\vec{E_j})  \nonumber \\ &&+\cdots, \label{eqn:three_terms}
\end{eqnarray}
where $Pr\{\vec{s_2} | \vec{s_1}\}$ indicates the transition probability of moving from state $\vec{s_1}$ to state $\vec{s_2}$. In (\ref{eqn:three_terms}), only the first three terms are related to the states that can move to the state $\vec{s}-\vec{E_i}+\vec{E_j}$  in only one step transition. The other terms are related to the states that can move to the state $\vec{s}-\vec{E_i}+\vec{E_j}$ in at least two step transitions. Since the transition probabilities of multi-step transitions are usually much smaller than those of one-step transitions, the other terms related to the states that require multi-step transitions to the state  $\vec{s}-\vec{E_i}+\vec{E_j}$ are ignored. The transition probabilities $Pr\{\vec{s}-\vec{E_i}+\vec{E_j}| \vec{s}-\vec{E_i}\}$,  $Pr\{\vec{s}-\vec{E_i}+\vec{E_j}| \vec{s}\}$, and  $Pr\{\vec{s}-\vec{E_i}+\vec{E_j} | \vec{s}+\vec{E_j}\}$ can be computed as follows:
\begin{eqnarray}
 &&Pr\big\{\vec{s}-\vec{E_i}+\vec{E_j}\big | \vec{s}-\vec{E_i}\big\} =  \frac{\lambda_j}{R(\vec{s}-\vec{E_i})},  \\
&& Pr\big\{\vec{s}-\vec{E_i}+\vec{E_j}\big | \vec{s}\big\} = \frac{n_i v_{ij}}{R(\vec{s})},\\  
&& Pr\big\{\vec{s}- \vec{E_i}+\vec{E_j}\big | \vec{s}+\vec{E_j}\big\}= \frac{n_i\mu}{R(\vec{s}+\vec{E_j})}, 
\end{eqnarray}
where $R(\vec{s})$ represents the total departure rate of state $\vec{s}=(n_1, n_2, \cdots, n_M)$, and it can be approximated as
\begin{equation}
 R(\vec{s}) \approx \lambda+\sum_{k=1}^M n_k\Big(\mu+\sum_{l\neq k} v_{kl}\Big). \label{eqn:a}
\end{equation}
Furthermore, if $p(\vec{s}-\vec{E_i}+\vec{E_j})$ is approximated as the sum of only these three terms present in (\ref{eqn:three_terms}), then the last term in (\ref{eqn:gg1}) can be approximated as
\begin{eqnarray}
&&\sum_{i=1}^M\sum_{j\neq i} (n_j+1) v_{ji} p(\vec{s}-\vec{E_i}+\vec{E_j}) I(n_i)  \nonumber \\
&\approx&\sum_{i=1}^M \sum_{j\neq i}\frac{(n_j+1) v_{ji}\lambda_j}{R(\vec{s}-\vec{E_i})}\cdot p(\vec{s}-\vec{E_i}) I(n_i)   \nonumber \\  &&+\sum_{i=1}^M \sum_{j\neq i}\frac{ n_iv_{ij}(n_j+1) v_{ji}}{R(\vec{s})}\cdot p(\vec{s}) \nonumber \\  &&+ \sum_{i=1}^M \sum_{j\neq i}\frac{ n_i\mu(n_j+1) v_{ji}}{R(\vec{s}+\vec{E_j})}\cdot p(\vec{s}+\vec{E_j}).  \label{eqn:ggv}
\end{eqnarray}
Notably, the last term in (\ref{eqn:ggv}) can be rewritten as follows:
\begin{eqnarray}
&& \sum_{i=1}^M \sum_{j\neq i}\frac{ n_i\mu(n_j+1) v_{ji}}{R(\vec{s}+\vec{E_j})}\cdot p(\vec{s}+\vec{E_j}) \nonumber \\ 
&=&\sum_{j=1}^M \sum_{i\neq j}\frac{ n_i\mu(n_j+1) v_{ji}}{R(\vec{s}+\vec{E_j})}\cdot p(\vec{s}+\vec{E_j}) \nonumber \\ 
&=&\sum_{i=1}^M \frac{(n_i+1)\mu\sum_{j\neq i} n_j v_{ij}}{R(\vec{s}+\vec{E_i})}\cdot p(\vec{s}+\vec{E_i}). \label{eqn:v}
\end{eqnarray}
Therefore, the global balance equation (\ref{eqn:gg1}) can be rearranged as follows:
\begin{eqnarray}
p(\vec{s})\sum_{i=1}^M\Big\{n_i\Big[\mu +\sum_{j\neq i} v_{ij}\big(1-\frac{(n_j+1) v_{ji}}{R(\vec{s})}\big)\Big] I(n_i)+\lambda_i \Big\} && \nonumber \\
\approx\sum_{i=1}^M \Big[\lambda_i+\frac{\sum_{j\neq i}(n_j+1) v_{ji}\lambda_j}{R(\vec{s}-\vec{E_i})}\Big]\cdot p(\vec{s}-\vec{E_i}) I(n_i) \hspace*{0.5cm}&&  \nonumber \\  + \sum_{i=1}^M (n_i+1)\mu\Big[1+\frac{\sum_{j\neq i}n_j v_{ij}}{R(\vec{s}+\vec{E_i})}\Big]\cdot p(\vec{s}+\vec{E_i}). \hspace*{0.8cm}  && \label{eqn:ggg1}  
\end{eqnarray}
If both the following two local balance equations
\begin{eqnarray}
n_i\Big[\mu +\sum_{j\neq i} v_{ij}\Big(1-\frac{(n_j+1) v_{ji}}{R(\vec{s})}\Big)\Big] p(\vec{s}) && \nonumber \\
=\Big[\lambda_i+\frac{\sum_{j\neq i}(n_j+1) v_{ji}\lambda_j}{R(\vec{s}-\vec{E_i})}\Big] p(\vec{s}-\vec{E_i}) && \label{eqn:lll1}
\end{eqnarray}
and
\begin{equation}
 \lambda_i p(\vec{s})= (n_i+1)\mu\Big[1+\frac{\sum_{j\neq i}n_j v_{ij}}{R(\vec{s}+\vec{E_i})}\Big] p(\vec{s}+\vec{E_i}) \label{eqn:lll2}
\end{equation}
can hold simultaneously, then the global balance equation (\ref{eqn:ggg1}) holds. Unfortunately, Eqs. (\ref{eqn:lll1}) and (\ref{eqn:lll2}) are not consistent. However, the weighted sum of Eq. (\ref{eqn:lll1}) and Eq. (\ref{eqn:lll3}), which is a shift version of (\ref{eqn:lll2}),
\begin{equation}
n_i\mu\Big[1+\frac{\sum_{j\neq i}n_j v_{ij}}{R(\vec{s})}\Big]  p(\vec{s})= \lambda_i p(\vec{s}-\vec{E_i}), \label{eqn:lll3}
\end{equation}
can be used to form an approximate local balance equation for solving $p(\vec{s})$, as given by 
\begin{equation}
 p(\vec{s})= \frac{f_i(\vec{s})}{n_i} p(\vec{s}-\vec{E_i}) \label{eqn:approx_local}
\end{equation}
The function $f_i(\vec{s})=f_i(n_1,\cdots,n_M)$ in (\ref{eqn:approx_local}) is given by
\begin{eqnarray}
f_i(\vec{s}) = \alpha\times\frac{\lambda_i+\frac{\sum_{j\neq i}(n_j+1) v_{ji}\lambda_j}{R(\vec{s}-\vec{E_i})}}{\mu +\sum_{j\neq i} v_{ij}\big[1-\frac{(n_j+1) v_{ji}}{R(\vec{s})}\big]} && \nonumber \\ +(1-\alpha)\times\frac{\lambda_i}{\mu\Big[1+\frac{\sum_{j\neq i}n_j v_{ij}}{R(\vec{s})}\Big]}, \hspace*{0.3cm} && \label{eqn:f_i}
\end{eqnarray}
where $0\leq \alpha\leq 1$ is an average factor. Since the local balance equations in (\ref{eqn:lll1}) and (\ref{eqn:lll3}) are not consistent, the solution solved solely by (\ref{eqn:lll1}) or (\ref{eqn:lll3}) is either overestimated or underestimated. The parameter $\alpha$ is just used to balance these two solutions to obtain a more accurate solution. In (\ref{eqn:f_i}), when $v_{ij}=0$ for any $i$ and $j$, $f_i(\vec{s})$ reduces to $\lambda_i/\mu$. Using (\ref{eqn:approx_local}), one can derive a similar formula as (\ref{eqn:ps}) for the stationary state probability $p(\vec{s})$ except that the parameter $\lambda_i/\mu$ in (\ref{eqn:ps}) must be replaced with $f_i(\vec{s})$ given  by (\ref{eqn:f_i}). Therefore, the blocking rate $P_b$ and the bandwidth usage $E(Y)$ can be calculated using the following two equations (\ref{eqn:blocking1}) and (\ref{eqn:EY1}), respectively.
\begin{equation}
\hspace*{-0.3cm}P_b=\frac{\displaystyle\sum_{m=1}^M \sigma_m \sum_{\ \forall (n_1,\cdots,n_M)\in B_m } \prod_{i=1}^M \frac{[f_i(n_1,\cdots,n_M)]^{n_i}}{n_i!}}{\displaystyle \sum_{\forall (n_1,\cdots,n_M)\in S\ } \prod_{i=1}^M \frac{[f_i(n_1,\cdots,n_M)]^{n_i}}{n_i!}}. \label{eqn:blocking1}
\end{equation}
\begin{equation}
\hspace*{-0.3cm}E(Y)=\frac{\displaystyle\sum_{\forall (n_1,\cdots,n_M)\in S}\Bigg(\sum_{i=1}^M n_ic_i\Bigg) \prod_{i=1}^M\frac{[f_i(n_1,\cdots,n_M)]^{n_i}}{n_i!}}{\displaystyle\sum_{\forall (n_1,\cdots,n_M)\in S\ } \prod_{i=1}^M\frac{[f_i(n_1,\cdots,n_M)]^{n_i}}{n_i!}}. \label{eqn:EY1}
\end{equation}

Since the required bandwidth resources for a connection is not similar under different MCSs, an ongoing connection may be dropped due to lack of bandwidth resources when the connection changes its MCS. Thus,  the dropping rate of unicast IPTV connections, resulting from the change of MCS or user mobility,  is analyzed in the following. Define $D_{mj}$ to be the set of all states $\vec{s}$ that satisfy $Y>K-c_j+c_m$ and $n_m>0$. That is, if the system is in the state $\vec{s}\in  D_{mj}$, $j<m$, then the connection using MCS $m$ must be dropped when the connection changes its MCS rate from $r_m$ to $r_j$, owing to the lack of bandwidth resources. Under the mobility model defined in this paper, only when the system is in the state $\vec{s}\in D_{m(m-1)}$, where $2\leq m\leq M$, may the connections using MCS $m$ be dropped. The case $m=1$ corresponds to the handover events that users move outward to the adjacent cells. Since the dropping rate of handover events depends on the states of adjacent cells,  it is not considered here. According to the definition of $D_{mj}$, the probability that the connection using MCS $m$ is dropped due to the change of MCS and lack of bandwidth resources is expressed by
\begin{equation}
\hspace*{-0.3cm}Pr\{{\rm dropped}\ |\ {\rm MCS}\ m\}= \frac{\displaystyle\sum_{j<m}\Big(\frac{v_{mj}}{\sum_{i\neq m}v_{mi}}\Big)\sum_{ \forall \vec{s}\in D_{mj}}p(\vec{s})}{\displaystyle\sum_{\forall \vec{s}\in S, n_m>0}p(\vec{s})}. \label{eqn:dropping}
\end{equation}

Notably, only when the connection switches to a lower MCS rate may it be dropped. Hence, the term $v_{mj}/\sum_{i\neq m}v_{mi}$, $j<m$, must be multiplied in (\ref{eqn:dropping}). For the mobility model defined in this paper,  $v_{mj}/\sum_{i\neq m}v_{mi}=1/2$ if $j=m-1$ and $v_{mj}/\sum_{i\neq m}v_{mi}=0$, otherwise. Subsequently, the mean dropping rate $P_d$ in the cell (excluding that of handover events) can be calculated by
\begin{equation}
\hspace*{-0.3cm}P_d=\sum_{m=2}^M \sigma^\prime_m Pr\{{\rm dropped}\ |\ {\rm MCS}\ m\}, \label{eqn:dropping1}
\end{equation}
where $\sigma^\prime_m$ is the ratio of the number of connections using MCS $m$ to the total number of connections using MCSs 2 to $M$ and can be obtained using the flow conservation law. Under the mobility model defined in this work, $ \sigma^\prime_m=1/(M-1)$ since the mobility model with similar transition rate in both moving directions makes the numbers of connections in individual zones equal in the steady state according to the flow conservation law.

\section{Numerical Results}
In this section, the blocking rate, dropping rate, and bandwidth usage of unicast IPTV services with AMC are evaluated based on the analytical results in Section III. Simulation results with 99\% confidence intervals are also given for verification.  In simulations, the arrival process of unicast IPTV requests follows the Poisson process with total arrival rate $\lambda$. The total amount of bandwidth resources reserved for unicast IPTV services, $K$, equals 20 if not mentioned. The watching time $T$ of each request is exponentially distributed with average $1/\mu=20$ minutes, namely $\mu$= 0.05 requests/min if not specifically mentioned. The sojourn time $W$ of a user staying in each zone of a cell is also assumed to be exponentially distributed with average time $w$ minutes. The transition rates between zones are assumed to be homogeneous and each user in Zone $m$ can move to Zone $m-1$ or Zone $m+1$ equally probably.  That is, the transition rate $v_{ij}$ is set to be $v$ for all $|i-j|=1$. Therefore, the parameter $v$, the  transition rate of changing MCS rate from $r_i$ to $r_j$ during a session, equals $\frac{1}{2}\times Pr\{W<T\}\times \frac{1}{w}$. Since $Pr\{W<T\}=(1/w)/(\mu+1/w)$,  $v$ can be computed as follows:
\begin{equation}
v=\frac{1}{2}\times\frac{1/w}{\mu+1/w}\times \frac{1}{w}. \label{eqn:w}
\end{equation}
Notably, the users in Zone 1 may move outward to adjacent cells (handover events) and leave the considered system (cell) with the handover rate $v$. 

\begin{table}
\caption{Required resource consumption of each unicast connection under different MCS rates \cite{5}.}
\centering
\begin{tabular}{ccccc}
\hline  & & Required   & Required Bandwidth& Area \\
Zone & MCS & Slots/Frame &  Resources/Connection  & Fraction\\
\hline 1 & QPSK 1/2 & 14  & 1 & 0.166352 \\
 2 & QPSK 3/4 & 9  & 9/14 & 0.287335 \\
 3 & 16-QAM 1/2  & 7 & 7/14 & 0.120983 \\
 4 & 16-QAM 3/4 & 5  & 5/14 & 0.236295 \\
 5 & 64-QAM 2/3 & 4 & 4/14 & 0.068053 \\
6 & 64-QAM 3/4 & 3 & 3/14 & 0.120983 \\
\hline
\end{tabular}\label{table:MCS}
\end{table}

\begin{figure}
\centering
\includegraphics[width=3.5in]{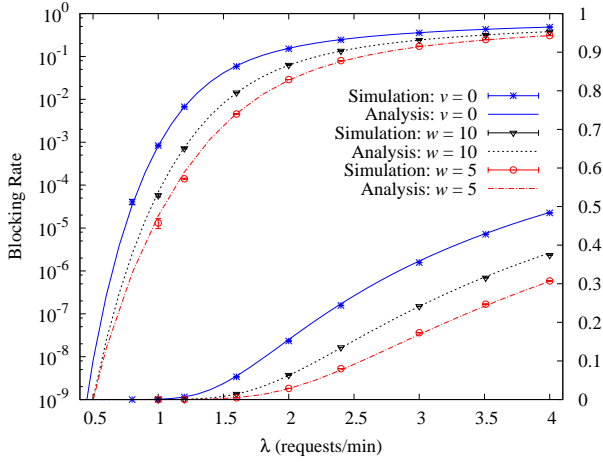}
\caption{Blocking rates of unicast IPTV services with AMC.} \label{fig2}
\end{figure}

\begin{figure}
\centering
\includegraphics[width=3.5in]{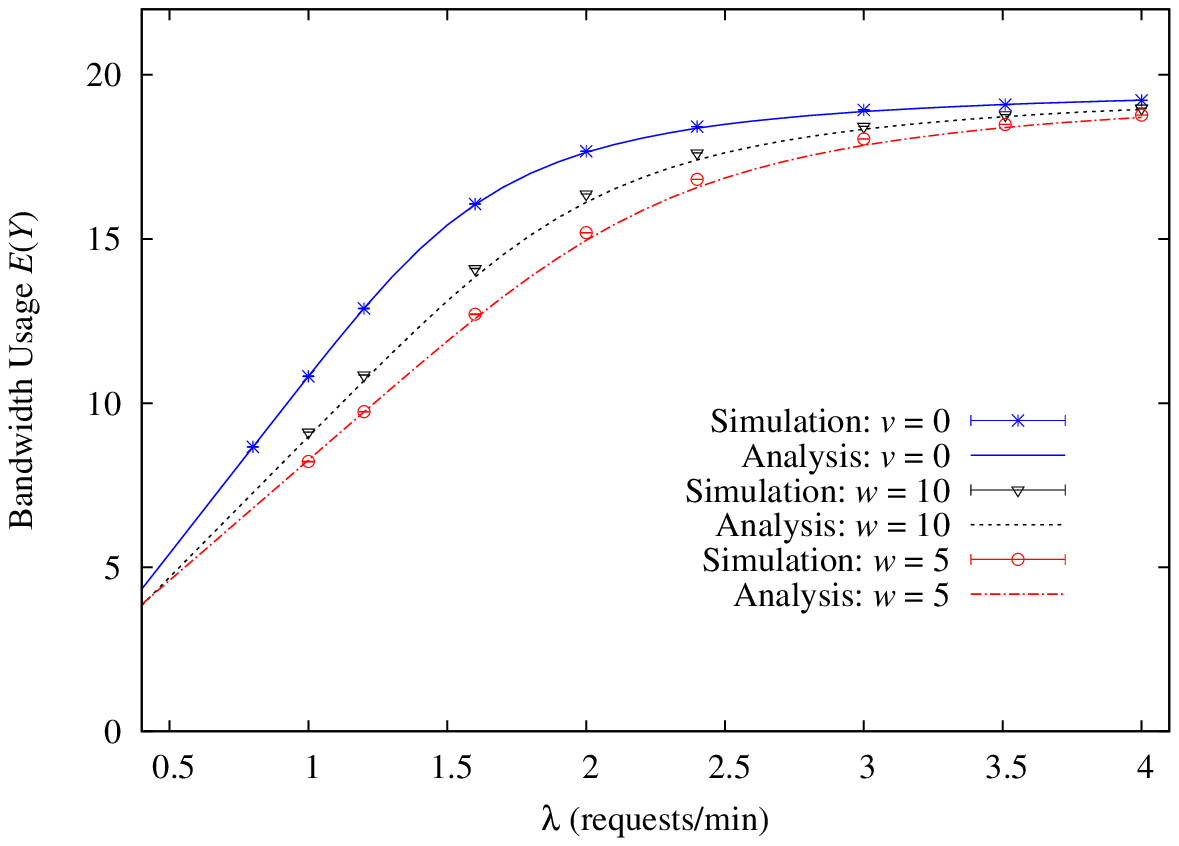}
\caption{Bandwidth usage of unicast IPTV services with AMC.}\label{fig3}
\includegraphics[width=3.5in]{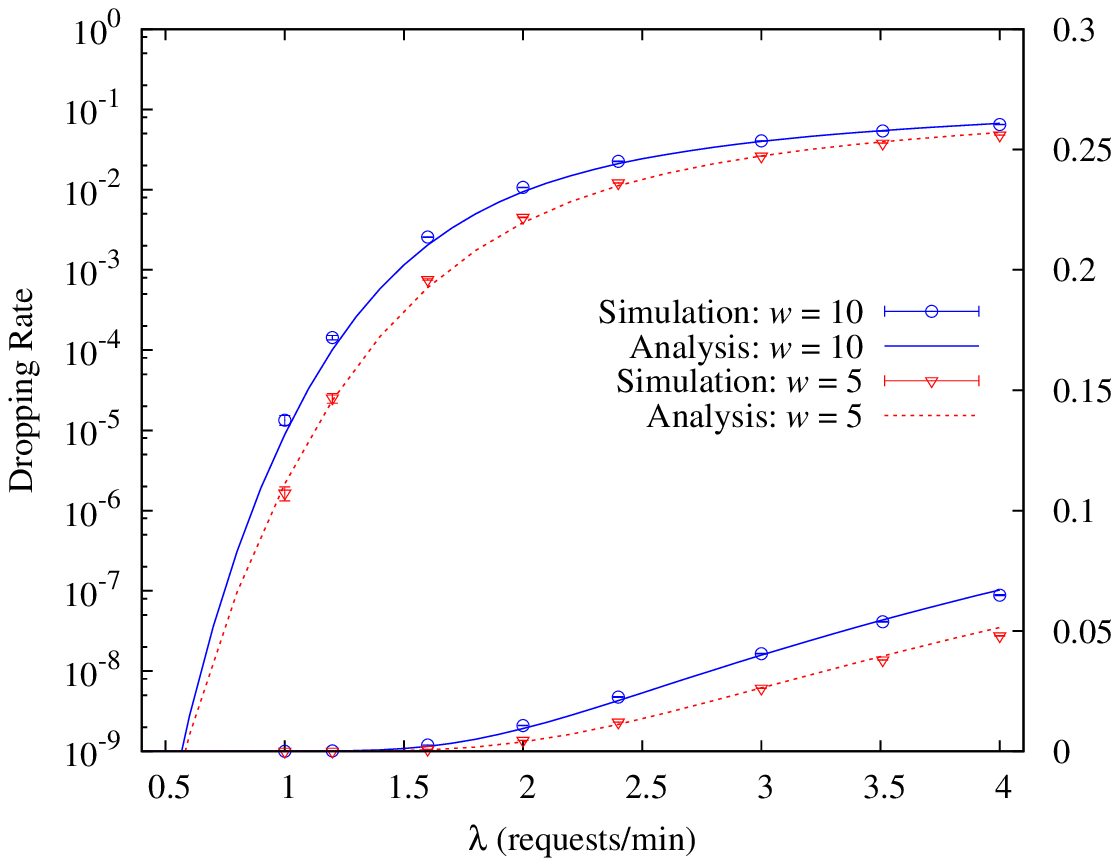}
\caption{Dropping rates of unicast IPTV services under mobility model.} \label{fig:dropping}
\end{figure}

In the first example, six MCSs  with different amounts of required bandwidth resources and area fractions which are calculated in \cite{5} are considered and listed in Table \ref{table:MCS}. Although the used parameters are coming from the WiMAX-based network, the proposed solutions can be directly applied to LTE-A 4G or 5G networks. The required amount of bandwidth resources of each unicast IPTV connection is assumed to be one unit when the MCS scheme QPSK 1/2 is used. Figure \ref{fig2} shows the blocking rate performance, plotted in both logarithmic and normal scales, under different values of $\lambda$. Three cases, including $v=0$, $w=5$, and 10 minutes, are investigated in Fig. \ref{fig2}. The case $v=0$ is the scenario that all unicast connections do not change their MCS rates during their sessions. The blocking rate for the case $v=0$ is computed based on Eq. (\ref{eqn:blocking}). The cases $w=10$ and $w=5$ minutes belong to the scenario that users maybe move to adjacent zones after staying in a zone for an exponentially distributed random time of average $w$ minutes. The blocking rates for the cases $w=10$ and $w=5$ minutes are calculated based on Eq. (\ref{eqn:blocking1}), where the average factor $\alpha$ present in (\ref{eqn:f_i}) is set to 0.4. Numerical results in Fig. \ref{fig2} demonstrate that analytical results are exactly coincident with simulation results for the case $v=0$. As for the cases $w=10$ and 5 minutes, the errors between analysis and simulation results  are negligible. Figure \ref{fig3} indicates the corresponding bandwidth usages of unicast IPTV services for these three cases mentioned above. The bandwidth usage is calculated based on Eq. (\ref{eqn:EY}) for the case $v=0$ or Eq. (\ref{eqn:EY1}) for the cases $w=10$ and $w=5$ minutes.  The corresponding dropping rate of the system with user mobility, calculated based on (\ref{eqn:dropping1}), is plotted in Fig. \ref{fig:dropping}.
Numerical results in Figs.  \ref{fig3} and \ref{fig:dropping} also reveal that analytical results are very accurate. In Figs.  \ref{fig2} and  \ref{fig3}, the blocking rate and the bandwidth usage of the system with mobility are smaller than those without mobility. This is because in the system with user mobility, mobile users in Zone 1 may move outward to adjacent cells and leave the considered system (cell), yielding the reduction of the system load. Additionally, some ongoing connections may be dropped when they change their MCSs, yielding the decrease of the system load as well. Thus, the blocking rate and bandwidth usage of the system with mobility are smaller than those without mobility. Similarly, the decrease of the load in the system with mobility parameter $w=5$ minutes is larger than that with $w=10$ minutes. Hence, the blocking rate, dropping rate, and bandwidth usage of the system with $w=5$ minutes are respectively less than those with $w=10$ minutes. However, the results in Figs.  \ref{fig2} to  \ref{fig:dropping} do not imply that the user mobility improves the system performance of a wireless cellular network. On the contrary, there exist more challenging problems such as the handover and connection dropping in the system with user mobility.

\begin{figure}
\centering
\includegraphics[width=3.5in]{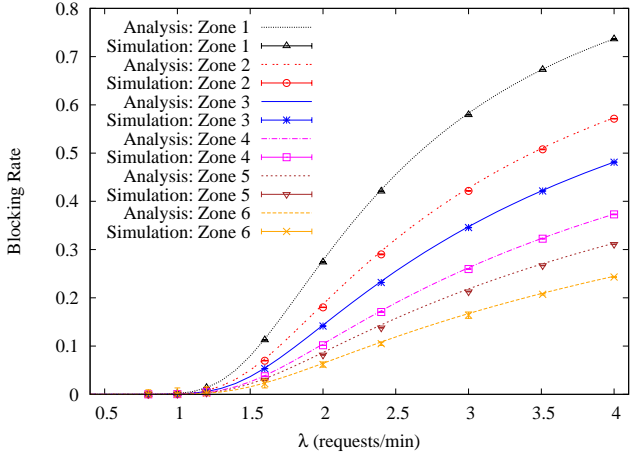}
\caption{Blocking rates of unicast IPTV services with AMC in individual zones.} \label{fig4}
\end{figure}

Continuing the example in Fig. \ref{fig2}, the blocking rates of individual zones in a cell under the case $v=0$ are compared in Fig. \ref{fig4}. The blocking rate of each zone is computed based on Eq. (\ref{eqn:reject}). According to the results in Fig. \ref{fig4}, the blocking rate of Zone 1 is the highest  while that of Zone 6 is the lowest.  This is because the required amount of bandwidth resources per connection is higher when a lower MCS rate is used, yielding a larger blocking rate. Therefore, the AMC scheme favors the mobile users who are closer to the base station. Figure \ref{fig4} again demonstrates that the presented analytical results are exactly coincident with simulation results.

\begin{figure}
\centering
\includegraphics[width=3.5in]{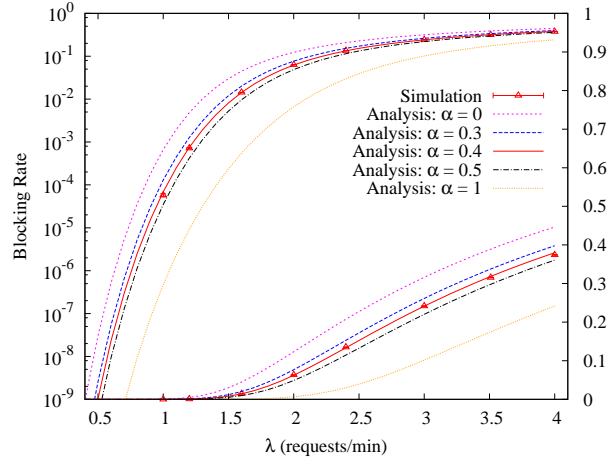}
\caption{Estimated blocking rates using different values of $\alpha$ for scenario with change of MCS. ($w=10$ min)} \label{fig5}
\end{figure}

\begin{figure}
\centering
\includegraphics[width=3.5in]{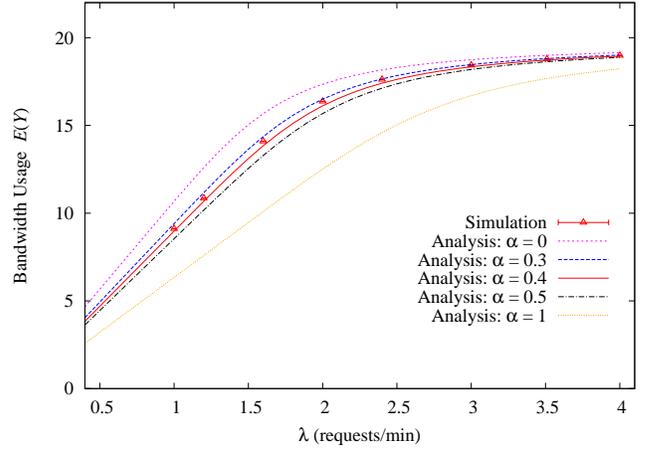}
\caption{Estimated bandwidth usage using different values of $\alpha$ for scenario with change of MCS. ($w=10$ min)}\label{fig6}
\end{figure}

According to Eq. (\ref{eqn:f_i}), the accuracy of the proposed analytical results is related to the average factor $\alpha$. Therefore, the impact of $\alpha$ on the accuracy of analytical results is studied. The parameter $w$ is set to 10 minutes and all MCS schemes listed in Table \ref{table:MCS} are used. Figure  \ref{fig5} indicates the estimated blocking rate performance, including both the logarithmic and normal scales, under different values of $\alpha$. Figure \ref{fig6} shows the corresponding estimated bandwidth usage performance. According to the results, the accuracy of the analysis is excellent if the average factor $\alpha$ is set to 0.4. Moreover,  if $\alpha=0$, then the analysis results based on (\ref{eqn:blocking1}) and  (\ref{eqn:EY1}) become upper bounds of the blocking rate and the bandwidth usage, respectively. While if $\alpha=1$, the analysis results become lower bounds of the blocking rate and the bandwidth usage. Therefore, the approximate solution for $p(\vec{s})$ presented in Section III.B can be accurate by setting a proper value for $\alpha$. 


To find the proper values of $\alpha$, a more detailed study on $\alpha$ must be done. Hence, several experiments under different scenarios are conducted to find the proper values of $\alpha$. According to Eq. (\ref{eqn:f_i}), $f_i (\vec{s})$ is mainly affected by the values of  $\mu$ and $\sum_{j\neq i}v_{ij}$ because $(n_j+1)v_{ji}/R(\vec{s})$ and $\sum_{j\neq i}n_jv_{ij}/R(\vec{s })$ are usually much less than 1. Therefore, the proper value of  $\alpha$ mainly depends on the value $\mu/\sum_{j\neq i}v_{ij}$. From Eq. (\ref{eqn:w}),  one can derive that $\sum_{j\neq i}v_{ij}=2v=(1/w)^2/(\mu+1/w)$. It follows that $\mu/\sum_{j\neq i}v_{ij}=(\mu w)^2+\mu w$. That is, the choice of $\alpha$ mainly depends on the value of $\mu w$. Consequently, several experiments are conducted to find the proper values of $\alpha$ under different values of $\mu w$. The proper values of $\alpha$ are found by trial and error and the results are shown in Table \ref{table:alpha}. From Table \ref{table:alpha}, the proper value of $\alpha$ is not a linear function of $\mu w$. Since $\mu/\sum_{j\neq i}v_{ij}=(\mu w)^2+\mu w$ is a polynomial of $\mu w$ with order 2,  the parabolic curve is used to fit the behavior of $\alpha$. The fitting function is obtained using the non-linear regression approach and it is given by $\alpha = 1.48x^2-1.22x+0.63$, where $x=\mu w$. Since $0\leq \alpha\leq 1$, the fitting function is expressed by
\begin{equation}
\alpha=\min\{1.48x^2-1.22x+0.63, 1\}, \ \ \ x=\mu w.
\label{eqn:alpha}
\end{equation}

\begin{table}
\caption{Some proper values of $\alpha$ under different scenarios. ($K=20$, $M=6$)}
\centering
\begin{tabular}{ccc}
\hline  ($1/\mu$, $w$) min & $\mu w$ & $\alpha$ \\
\hline (60, 5) & 0.083  & 0.55\\
(5, 0.5) & 0.100  & 0.52\\
(25, 5) & 0.200  & 0.43\\
(20, 5) & 0.250  & 0.40\\
(20, 8) & 0.400  & 0.38\\
(10, 5), (20, 10) & 0.500  & 0.40\\
(7, 5), (14, 10) & 0.714  & 0.50\\
(6, 5), (12, 10) & 0.833  & 0.65\\
(5, 5), (10, 10) & 1.000  & 0.90\\
(4.5, 5), (9, 10) & 1.111  & 1.00\\
\hline
\end{tabular}\label{table:alpha}
\end{table}

\begin{figure}
\centering
\includegraphics[width=3.5in]{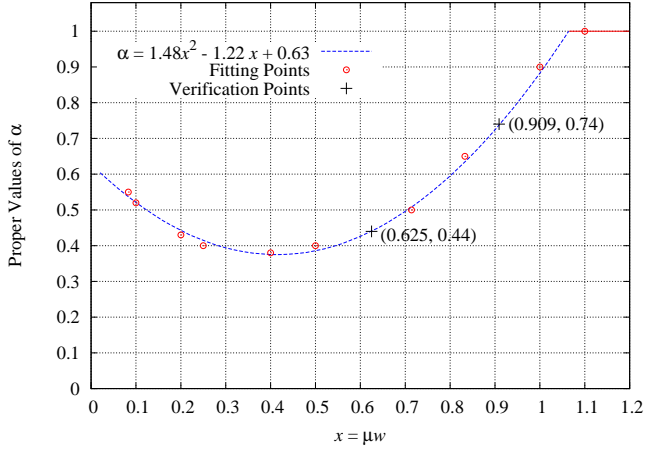}
\caption{The fitting curve for the proper value of $\alpha$.}\label{fig:alpha}
\end{figure}

\begin{figure}
\centering
\includegraphics[width=3.5in]{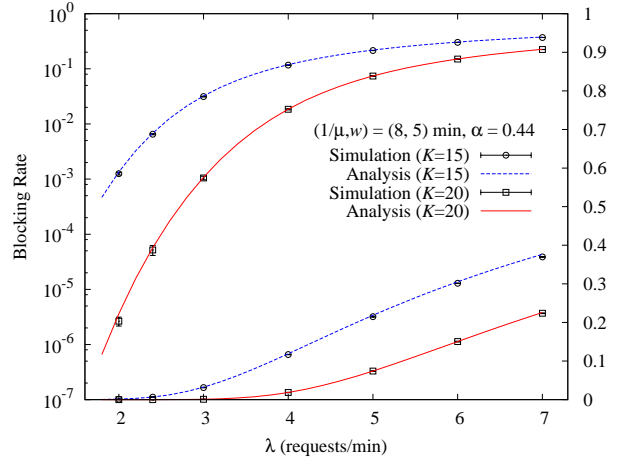}
\caption{Comparison between simulation and analysis results under different values of $K$. ($\alpha=0.44, (1/\mu, w) = (8, 5)$ min)}\label{fig:check1}
\end{figure}

\begin{figure}
\centering
\includegraphics[width=3.5in]{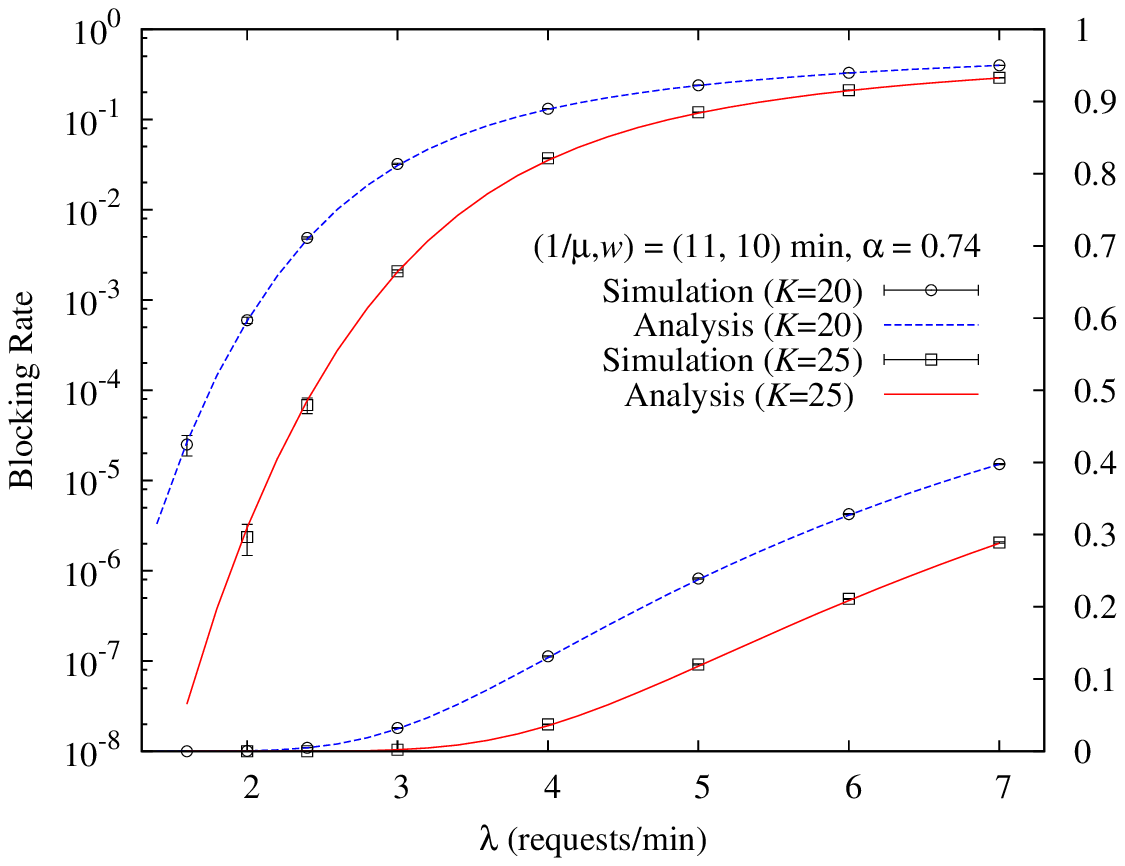}
\caption{Comparison between simulation and analysis results under different values of $K$. ($\alpha=0.74, (1/\mu, w) = (11, 10)$ min)}\label{fig:check2}
\end{figure}

Figure \ref{fig:alpha} shows the fitting curve for $\alpha$. To validate the correctness of the fitting function, two points ($\mu w$, $\alpha$) = (0.625, 0.44) and (0.909, 0.74) in Fig. \ref{fig:alpha} are checked and verified. For the case ($\mu w$, $\alpha$) = (0.625, 0.44), the parameters $1/\mu$ = 8 min, $w$ = 5 min, and $\alpha$ = 0.44 are used. For the case ($\mu w$, $\alpha$) = (0909, 0.74), the parameters $1/\mu$ = 11 min, $w$ = 10 min, and $\alpha$ = 0.74 are used. The other parameters not mentioned are the same as those in Fig. \ref{fig2}. The  simulation and analysis results for these two cases are shown in Figs. \ref{fig:check1} and \ref{fig:check2}, respectively. Figures \ref{fig:check1} and \ref{fig:check2} demonstrate that the analysis results are coincident with the simulation results under different values of $K$. Hence, using the proposed fitting function for $\alpha$, one can easily calculate the proper values of $\alpha$ under different scenarios.

\begin{table}
\caption{Required resource consumption of each unicast connection when two MCS schemes are used.}
\centering
\begin{tabular}{ccccc}
\hline  & & Required   & Required Bandwidth& Area \\
Zone & MCS & Slots/Frame &  Resources/Connection  & Fraction\\
\hline 1 & QPSK 1/2 & 14  & 1 & 0.453687 \\
 2 & 16-QAM 1/2 & 7  & 7/14 & 0.546313 \\
\hline
\end{tabular}\label{table:2MCS}
\end{table}
\begin{table}
\caption{Required resource consumption of each unicast connection when three MCS schemes are used.}
\centering
\begin{tabular}{ccccc}
\hline  & & Required   & Required Bandwidth& Area \\
Zone & MCS & Slots/Frame &  Resources/Connection  & Fraction\\
\hline 1 & QPSK 1/2 & 14  & 1 & 0.166352 \\
 2 & QPSK 3/4 & 9  & 9/14 & 0.408318 \\
 3 & 16-QAM 3/4 & 5  & 5/14 & 0.425331 \\
\hline
\end{tabular}\label{table:3MCS}
\end{table}
\begin{figure}
\centering
\includegraphics[width=3.5in]{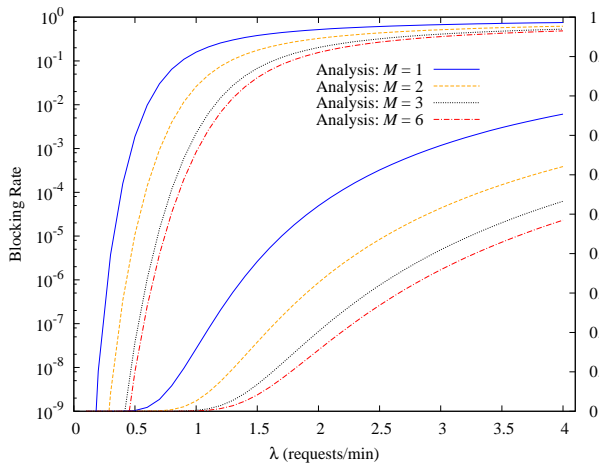}
\caption{Blocking rate performance under different numbers of MCSs.} \label{fig7}
\end{figure}
Subsequently, the relationship between system performance and the number of MCSs used is studied. The blocking rate and the bandwidth usage under the scenarios using a single MCS ($M=1$), two MCSs ($M=2$), three MCSs ($M=3$), and six MCSs ($M=6$) are compared. In the scenario using a single MCS, the only MCS QPSK 1/2 is used. In the scenario using two MCSs, QPSK 1/2 and 16-QAM 1/2 are employed, as shown in Table \ref{table:2MCS}. In the scenario using three MCSs, the MCSs listed in Table \ref{table:3MCS} are considered. In the scenario using six MCSs, the MCSs listed in Table  \ref{table:MCS} are employed.  The blocking rate is computed according to Eq. (\ref{eqn:blocking}), and the bandwidth usage is calculated based on Eq. (\ref{eqn:EY}). Figure \ref{fig7} depicts the blocking rates, including both the  logarithmic and normal scales, of various scenarios with $M=1$, 2, 3, and 6. The corresponding bandwidth usage performance is shown in Fig. \ref{fig8}. 
According to the results, the blocking rate and the bandwidth usage are significantly improved when three or six MCSs are used. Hence, the system performance of a wireless cellular network can be much improved if AMC is used in offering unicast IPTV services. Furthermore, the more the number of MCS schemes is used, the better the performance achieves. However, when the number of MCS schemes used is more than 3, the performance improvement relative to the case $M=3$ becomes slight. Additionally, the signaling overhead of AMC scheme, mainly resulting from the CQI feedback messages, increases with the number of MCSs used. Hence, there exists a trade-off problem between the system performance and the signaling overhead. Although AMC can improve the spectral efficiency, the blocking rate of unicast IPTV becomes unacceptably high when the load is large, as shown in Fig. \ref{fig2} or \ref{fig7}.  Accordingly, to further improve the system performance and enhance the spectral efficiency, hybrid unicast and multicast/broadcast schemes \cite{3,NVoD} can be considered for offering mobile IPTV services.
\begin{figure}
\centering
\includegraphics[width=3.5in]{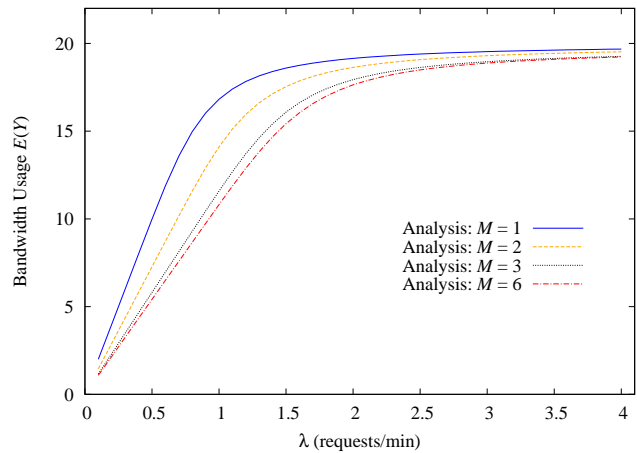}
\caption{Bandwidth usage performance under different numbers of MCSs.}\label{fig8}
\end{figure}

\begin{figure}
\centering
\includegraphics[width=3.5in]{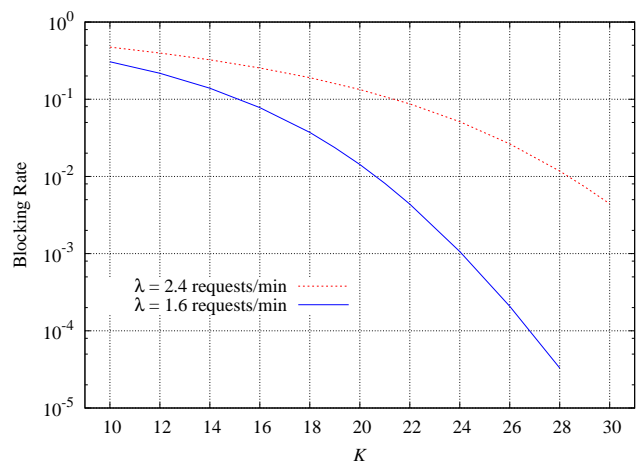}
\caption{Blocking rate versus $K$. ($M=6$, $1/\mu=20$ min, $w=10$ min)}\label{fig9}
\end{figure}

The presented analytical results can be applied to implement an accurate resource allocation for offering unicast IPTV services over a wireless cellular network. For example, given the request arrival rate $\lambda$, the appropriate amount of bandwidth resources $K$ to yield a blocking rate less than 1\% can be determined based on Eq. (\ref{eqn:blocking1}). Figure \ref{fig9} depicts the blocking rate performance versus $K$, the amount of allocated bandwidth resources for unicast IPTV services. In Fig. \ref{fig9}, to guarantee the blocking rate to be less than 1\%, the required amount of bandwidth resources $K$ must be larger than 21 if $\lambda=1.6$ requests/min and 29 if $\lambda=2.4$ requests/min. Since the arrival rate of IPTV users may vary significantly during a day, the bandwidth resources for unicast IPTV services must be dynamically allocated to maintain the system performance. According to the analytical results, network providers can calculate the blocking rates under different values of $\mu$, $\lambda$, $w$, and $K$, and store these results for later looking up. Network providers can monitor related parameters of IPTV services in real time and check whether the blocking rate requirement is satisfied or not. If the blocking rate is not satisfied, network providers must look up a proper $K$ from the calculated results and immediately allocate additional bandwidth resources for IPTV services. On the contrary, if the load of IPTV services is low and the allocated bandwidth resources is extra enough, then some bandwidth resources can be released for other services. 

Finally, to verify the usefulness of the proposed queueing model, a more realistic simulation model is conducted. The Markovian mobility model is replaced by the random walk mobility model \cite{random_walk} in simulations. In the random walk mobility model, a user randomly chooses a direction and travels straight over a fixed distance $d$ at  a constant velocity $f$ and then changes the moving direction $\theta$ that is uniformly distributed over the interval [0, 2$\pi$). The moving process is repeated continuously until the user leaves the system. The case of three zones with MCS schemes and area fractions shown in Table \ref{table:mobility} is considered. The parameters in the random walk mobility model are set to $d=0.08$ and $f=1.1\times 10^{-3}$/s, relative to the cell radius. If the cell radius is equal to 1000m, then the parameters in the random walk mobility model are $d=80$m and $f=4$ km/hr. Under the random walk mobility model, the transition rates between adjacent zones, as shown in Table \ref{table:transition}, are obtained via simulation and found to be not homogeneous. In Table \ref{table:transition}, the transition ``1 $\rightarrow$ 0" indicates the handover event. Using these transition rates, the blocking rate, bandwidth usage, and dropping rate are computed based on (\ref{eqn:blocking1}),  (\ref{eqn:EY1}), and (\ref{eqn:dropping1}), respectively. 
\begin{table}
\caption{MCS, resource consumption, and area fraction of each zone.}
\centering
\begin{tabular}{ccccc}
\hline  & & Required   & Required Bandwidth& Area \\
Zone & MCS & Slots/Frame &  Resources/Connection  & Fraction\\
\hline 1 & QPSK 1/2 & 14  & 1 & 0.453687 \\
 2 & 16-QAM 1/2 & 7  & 7/14 & 0.357278 \\
 3 & 64-QAM 2/3 & 4  & 4/14 & 0.189036 \\
\hline
\end{tabular}\label{table:mobility}
\end{table}
\begin{table}
\caption{Transition rates between zones under random walk mobility model.}
\centering
\begin{tabular}{|c|c|c|c|}
\hline  Direction & Transition Rate   & Direction & Transition Rate \\
\hline 3 $\rightarrow$ 2 & $v_{32}=$ 0.03576/min & 2 $\rightarrow$ 3  & $v_{23}=$ 0.1122/min \\
\hline 2 $\rightarrow$ 1  & $v_{21}=$ 0.1055/min &1 $\rightarrow$ 2  & $v_{12}=$ 0.1413/min  \\
\hline  1 $\rightarrow$ 0  &  $v_{10}=$ 0.1373/min  & 0 $\rightarrow$ 1  & X \\
\hline
\end{tabular}\label{table:transition}
\end{table}
\begin{figure}
\centering
\includegraphics[width=3.5in]{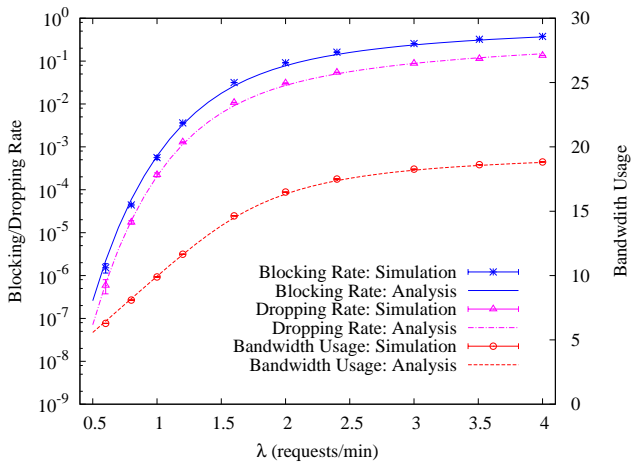}
\caption{Blocking rate, dropping rate, and bandwidth usage performance when the random walk mobility model is simulated. ($d=0.08$, $f=1.1\times 10^{-3}$/s)}\label{fig12}
\end{figure}
The average sojourn time $w$ of each user in a zone is about 11.9 minutes according to the simulation results. Since $\mu w$ = 0.595, the value of $\alpha$ is set to 0.42 according to the fitting function in (\ref{eqn:alpha}). To calculate the mean dropping rate of Zones 2 and 3, the parameter $\sigma^\prime_m$ in (\ref{eqn:dropping1}) must be determined based on the equalities $\sigma^\prime_2 v_{23}=\sigma^\prime_3 v_{32}$ (flow conservation law) and $\sigma^\prime_2+\sigma^\prime_3=1.$ That is, $\sigma^\prime_2 =v_{32}/(v_{23}+v_{32})$ and $\sigma^\prime_3=v_{23}/(v_{23}+v_{32}).$ Other parameters not mentioned here are the same as those in Fig. \ref{fig2}. Figure \ref{fig12} shows the simulation and analysis results, and demonstrates that the derived closed-form formulas can still accurately estimate the blocking rate, dropping rate, and bandwidth usage performance under the random walk mobility model.

\section{Conclusions}
The blocking rate, dropping rate, and bandwidth usage of unicast IPTV services using AMC over wireless cellular networks have been analyzed in this paper. In our analysis, exact closed-form solutions are derived  when the user mobility is ignored. Subsequently, approximate closed-form solutions are presented when the user mobility is considered. The analytical results are validated to be very accurate by simulations. According to the numerical results, the blocking rate and the bandwidth usage are much improved when AMC is used, indicating that AMC can effectively increase the spectral efficiency. The analytical results in this work can be employed by network providers for precise resource allocation and service planning in wireless cellular networks that support mobile IPTV services.  In addition, the derived analytical results can be applied to other real-time streaming applications such as voice and video conferencing services as well. 





\ifCLASSOPTIONcaptionsoff
  \newpage
\fi






\end{document}